# Pressure-induced Superconductivity in dual-topological semimetal Pt$_2$HgSe$_3$

Cuiying Pei[1#], Suhua Jin[1#], Peihao Huang[2#], Anna Vymazalova[4], Lingling Gao[1], Yi Zhao[1], Weizheng Cao[1], Changhua Li[1], Peter Nemes-Incze[5], Yulin Chen[1,6,7], Hanyu Liu[2,3*], Gang Li[1,6*], Yanpeng Qi[1*]

1. School of Physical Science and Technology, ShanghaiTech University, Shanghai 201210, China
2. State Key Laboratory of Superhard Materials and International Center for Computational Method and Software, College of Physics, Jilin University, Changchun 130012, China
3. International Center of Future Science, Jilin University, Changchun 130012, China
4. Czech Geological Survey, Prague 152 00, Czech Republic
5. Centre for Energy Research, Institute of Technical Physics and Materials Science, Budapest 1121, Hungary
6. ShanghaiTech Laboratory for Topological Physics, ShanghaiTech University, Shanghai 201210, China
7. Department of Physics, Clarendon Laboratory, University of Oxford, Parks Road, Oxford OX1 3PU, UK

\# These authors contributed to this work equally.
\* Correspondence should be addressed to Y.Q. (qiyp@shanghaitech.edu.cn) or G.L. (ligang@shanghaitech.edu.cn) or H.L(lhy@calypso.cn)

**ABSTRACT:**

Recently monolayer jacutingaite (Pt$_2$HgSe$_3$), a naturally occurring exfoliable mineral, discovered in Brazil in 2008, has been theoretically predicted as a candidate quantum spin Hall system with a 0.5 eV band gap, while the bulk form is one of only a few known dual-topological insulators which may host different surface states protected by symmetries. In this work, we systematically investigate both structure and electronic evolution of bulk Pt$_2$HgSe$_3$ under high pressure up to 96 GPa. The nontrivial topology persists up to the structural phase transition observed in the high-pressure regime. Interestingly, we found that this phase transition is accompanied by the appearance of superconductivity at around 55 GPa and the critical transition temperature $T_c$ increases with applied pressure. Our results demonstrate that Pt$_2$HgSe$_3$ with nontrivial topology of electronic states displays new ground states upon compression and raises potentials in application to the next-generation spintronic devices.

# INTRODUCTION

Quantum spin Hall insulators (QSHIs) constitute an important class of topological systems having a gapped insulating bulk and gapless helical edge states. Importantly, helical edge states, where the helical locking of spin and momentum suppresses backscattering of charge carriers, are considerable robust against interactions and nonmagnetic disorders, making QSHIs possess promising applications from low power electronics to quantum computing[1-8]. After the first experimental realization of a QSHI in the form of a HgTe/CdTe quantum well at cryogenic temperatures[4, 9], a quantum spin Hall state has subsequently been identified in exfoliated 1T′ phase of transition metal dichalcogenides (e.g. $WTe_2$) by scanning tunneling microscopy (STM)[10] and charge transport measurements[11]. Despite their massive fundamental interest and their prospective technological applications, a major challenge is the identification of large gap QSHI materials, which would enable room temperature dissipationless transport of their edge states.

Recently, a robust QSHI with a gap of up to 0.5 eV, which is one order of magnitude larger than that in $WTe_2$, has also been predicted in monolayer (ML) $Pt_2HgSe_3$[12]. The ternary compound $Pt_2HgSe_3$, so called Jacutingaite[13, 14], has a "sandwich-like" structure reminiscent of transition metal dichalcogenides, with a platinum layer between selenium and mercury. In the case of the monolayer, it was argued that the competition between large spin-orbit coupling, associated with Hg and Pt atoms, and sublattice symmetry breaking lead to a QSHI state robust at room temperature and switchable by external electric fields[12]. Furthermore, recent theoretical work found that bulk $Pt_2HgSe_3$ is one of only a few known dual-topological semimetals and may host different surface states protected by symmetries that are unrelated to the QSHI state[15-18]. However, to date only little is known from experiment about the bulk band structure supporting the different topological phases and topological surface state of $Pt_2HgSe_3$.

Specifically, Wu *et.al* predicted that the monolayer of $Pt_2HgSe_3$ to host different phases of unconventional superconductivity for finite hole and electron doping[19]. High pressure can effectively modify lattice structures and the corresponding electronic states in a systematic fashion. Indeed, superconductivity has been induced by the use of pressure in some topological compounds[20-27]. Here, we systematically investigate the high-pressure behavior of bulk $Pt_2HgSe_3$. Through *ab initio* band structure calculations, we find that the application of pressure does not qualitatively change the electronic and topological nature of the material until the structural phase transition is observed in the high-pressure regime. Interestingly, superconductivity appears beyond the structural phase transition and the maximum critical temperature, $T_c$, of 4.4 K at 88.8 GPa is observed. The results demonstrate that $Pt_2HgSe_3$ compounds with nontrivial topology of electronic states display new ground states upon compression.

**EXPERIMENTAL SECTION**

The high quality $Pt_2HgSe_3$ sample used in this work was synthesized from 3 individual elements by high-temperature solid state reactions[17].

An *in situ* high-pressure resistivity measurements were performed in a nonmagnetic diamond anvil cell (DAC). A cubic BN/epoxy mixture layer was inserted between BeCu gaskets and electrical leads. Four Pt foils were arranged in a van der Pauw four-probe configuration to contact the sample in the chamber for resistivity measurements. Pressure was determined by the ruby luminescence method[28]. An *in situ* high-pressure Raman spectroscopy investigation of $Pt_2HgSe_3$ was performed using a Raman spectrometer (Renishaw inVia, U.K.) with a laser excitation wavelength of 532 nm and low-wavenumber filter. A symmetric DAC with anvil culet sizes of 250 μm was used, with silicon oil as pressure transmitting medium (PTM). *In situ* high-pressure X-ray diffraction (XRD) measurements were performed at beamline BL15U of Shanghai Synchrotron Radiation Facility (X-ray wavelength λ = 0.6199 Å). Symmetric DACs with anvil culet sizes of 250 μm and Re gaskets were used. Silicon oil was used as the PTM and pressure was determined by the ruby luminescence method[28]. The two-dimensional diffraction images were analyzed using the FIT2D software[29]. Rietveld refinements on crystal structures under high pressure were performed using the General Structure Analysis System (GSAS) and the graphical user interface EXPGUI[30, 31].

The *ab initio* calculations were performed within the framework of density functional theory (DFT) by using QUANTUM ESPRESSO (QE) v6.5[32] with the exchange-correlation functional in form of generalized gradient approximation. The Perdew-Burke-Ernzerhof (PBE) pseudopotential files "pbe-n-kjpaw_psl.1.0.0" was used for Pt, Hg and Se. The kinetic energy cutoff with 40 Ry, a 8×8×8 $k$-mesh and a 4×4×4 $q$-mesh are used for the electron-phonon coupling calculations.

Structure prediction has been performed through a swam-intelligence-based CALYPSO method and its same-name code[33-35]. Density functional total energy calculations and structure relaxation are performed using the VASP plane-wave code[36,37]. We have adopted the PBE generalized gradient approximation density functional[38] and frozen-core all-electron projector-augmented wave (PAW) potentials[39] in our calculations. The electronic wave functions are expanded in a plane-wave basis set with a kinetic energy cutoff of 350 eV. The Brillouin zone (BZ) sampling is performed on k-meshes with a reciprocal space resolution of $2\pi \times 0.03$ Å$^{-1}$ to ensure that energies are converged to several meV/atom.

## RESULTS AND DISCUSSION

**Crystal structure and electronic properties of Pt$_2$HgSe$_3$**

Jacutingaite (Pt$_2$HgSe$_3$) is a layered platinum-group mineral, which has a centrosymmetric trigonal structure, belonging to the space group $P\bar{3}m1$ (No.164). As shown in Figure 1a-b, Hg atoms form a buckled honeycomb lattice surrounded by triangles of Pt and Se. There are two inequivalent platinum positions indicated by Pt1 and Pt2. The Pt1 atoms show an octahedral coordination with six selenium atoms, while Pt2 are surrounded by two mercury atoms in a trans position with respect to one another and four selenium atoms in a square planar coordination. The Pt1 and Pt2 octahedra are Se-Se edges shared and forms layers oriented parallel to (001), which is further AA-type stack along the $c$ axis. The $P\bar{3}m1$ phase is known to be topological at ambient pressure. In our *ab initio* calculations, we found that, at all pressured studied in this work, Pt$_2$HgSe$_3$ remains topological with the location of topological surface states being slightly modified. The details of the bulk, surface states and their evolutions under pressure will be discussed in later sections. Here, we show in Figure 1c and 1d the bulk and surface electronic structures at an arbitrary pressure 15.9 GPa. In Figure 1c, the dashed blue line and the solid red lines correspond to the bulk band structures without/with spin-orbital coupling. As is clearly shown, under pressure, $Pt_2HgSe_3$ remains metallic and topological, which holds as long as the space group is unchanged.

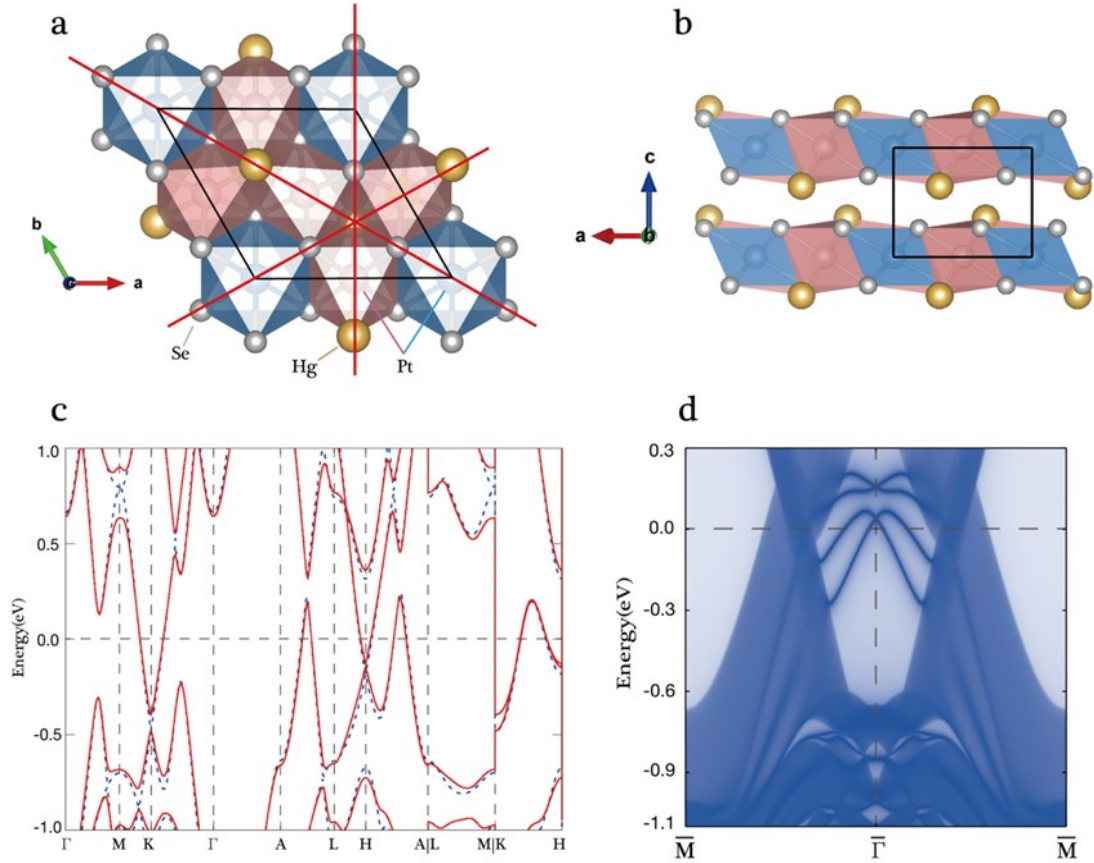

Figure 1 (a-b) Crystal structure of $Pt_2HgSe_3$. Pt1 and Pt2 octahedra which are shown in blue and red, respectively, denote two symmetry inequivalent local environment of Pt atom in the unit cell; (c) Electronic structure of $Pt_2HgSe_3$; (d) The corresponding states at (001) surface.

**Electrical resistivity at high pressure**

$Pt_2HgSe_3$ possess a typical layered structure, which is principally sensitive to external pressure. Hence, we measured $\rho(T)$ for $Pt_2HgSe_3$ using a non-magnetic DAC. Figure 2a shows the plots of temperature versus resistivity of $Pt_2HgSe_3$ for pressures up to 88.8 GPa. It reveals a metallic behavior in the whole pressure range. In a low-pressure region, increasing the pressure initially induces a weak but continuous suppression of the overall magnitude of $\rho$ with a minimum occurring at $P_{min} = 12$ GPa. Upon further

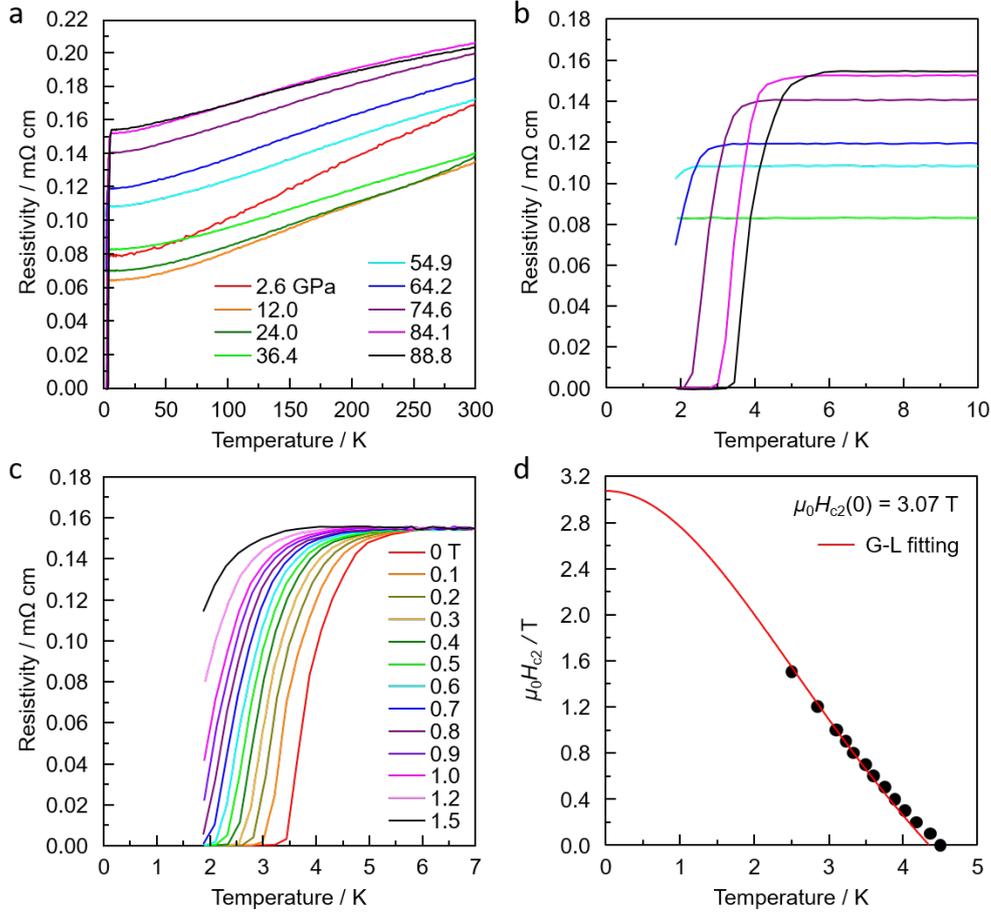

Figure 2 (a) Electrical resistivity of $Pt_2HgSe_3$ as a function of temperature for various pressures; (b) Temperature-dependent resistivity of $Pt_2HgSe_3$ in the vicinity of the superconducting transition; (c) Temperature dependence of resistivity under different magnetic fields for $Pt_2HgSe_3$ at 88.8 GPa; (d) Temperature dependence of upper critical field for $Pt_2HgSe_3$ at 88.8 GPa. Here, $T_c$ is determined as the 90% drop of the normal state resistivity. The solid lines represent the fits based on the Ginzburg–Landau (G-L) formula.

increasing pressure, the resistivity starts to increase gradually. At a pressure of 54.9 GPa, a small resistivity drop presents at 1.9 K, indicating superconducting phase transition. It should be noted that Mauro et.al has performed transport measurements on individual flakes of $Pt_2HgSe_3$ down to 250 mK and found no superconducting transition[40]. However, it cannot rule out that superconductivity appears in the material at pressures below 54.9 GPa, at temperatures lower then measured here. As shown in Figure 2b, the resistivity drop becomes more visible and the critical temperature $T_c$ increases to the maximum of 4.4 K at 88.8 GPa. The measurements on different samples of $Pt_2HgSe_3$ for three independent runs provide the consistent and reproducible results (Supplemental Information Fig. S1), confirming the intrinsic superconductivity under pressure. To gain insights into the superconducting transition, we applied the magnetic field for $Pt_2HgSe_3$ subjected to 88.8 GPa. When increasing $\mu_0H$, the resistivity drop is continuously shifted to a lower temperature (Figure 2c). The upper critical field, $\mu_0H_{c2}$, is determined using 90% point on the resistivity transition curves, and plots of $H_{c2}(T)$ are shown in Figure 2d. A simple estimate using the conventional one-band Werthamer-Helfand-Hohenberg approximation[41], neglecting the Pauli spin-paramagnetism effect and spin-orbit interaction, yields a value of 2.37 T for $Pt_2HgSe_3$. By using the Ginzburg–Landau (G-L) formula $\mu_0H_{c2}(T) = \mu_0H_{c2}(0) (1 − (T/T_c)^2)/(1 + (T/T_c)^2)$ to fit the data, the estimated $\mu_0H_{c2}(0)$ value is 3.07 T at 88.8 GPa. These fields are much lower than the Pauli limiting fields, $H_P(0) = 1.84T_c \sim 8.10$ T, respectively, indicating that Pauli pair breaking is not relevant.

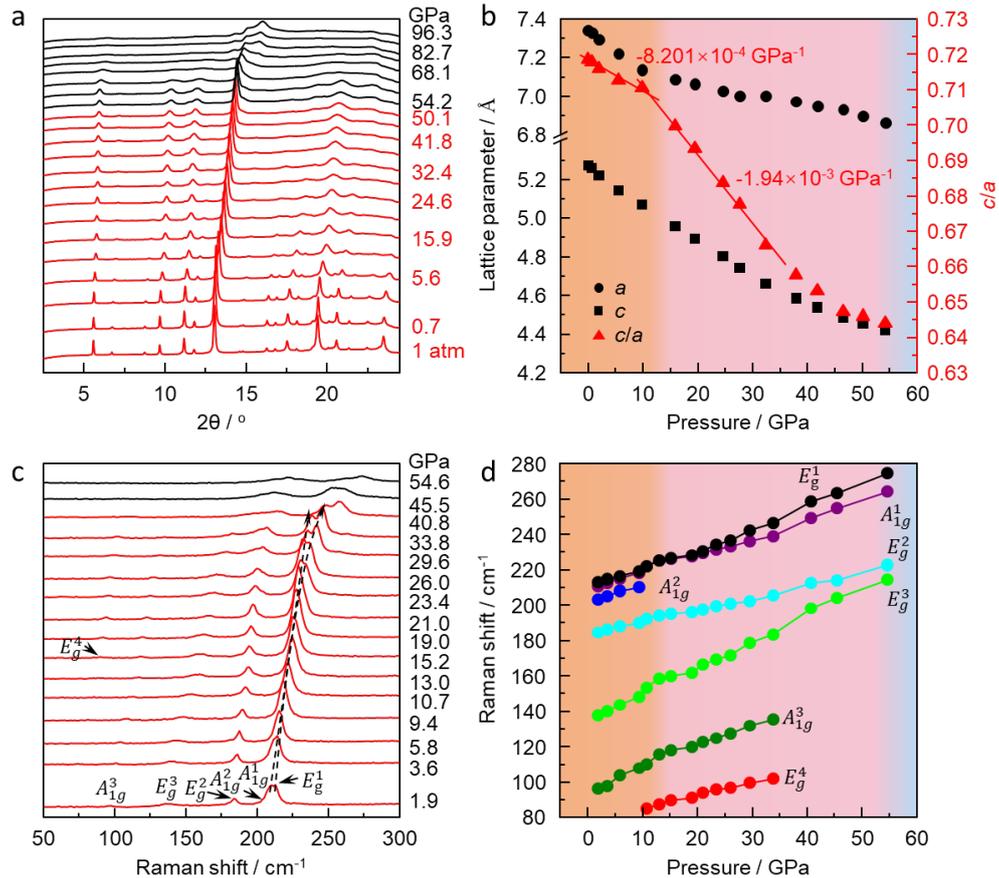

Figure 3. (a) XRD patterns of $Pt_2HgSe_3$ under pressure up to 96.3 GPa at room temperature with an x-ray wavelength of $\lambda = 0.6199$ Å. The red and black patterns distinguish phase transition with pressure over 50.1 GPa; (b) Pressure-dependence of lattice parameter $a$, $c$ and $c/a$ ratio for $Pt_2HgSe_3$ $P\bar{3}m1$ phase; (c) Raman spectra at various pressures for $Pt_2HgSe_3$; (d) Raman shift for $Pt_2HgSe_3$ in compression; the vibration modes display in increasing wavenumber order.

**Crystal structure evolution at high pressure**

To further identify the pressure-induced electron structure transition, *in situ* XRD measurements have been performed on Pt$_2$HgSe$_3$ to analysis the structure evolution under various pressures. Figure 3a displays the high-pressure synchrotron XRD patterns of Pt$_2$HgSe$_3$ measured at room temperature up to 96.3 GPa. A representative refinement at 1 atm is displayed in Figure S2. All the diffraction peaks can be indexed well to a trigonal structure with space group $P\bar{3}m1$ based on Rietveld refinement with GSAS software package.[30] As shown in Figure 3b both *a*-axis and *c*-axis lattice constants decrease with increasing pressure. The structure of Pt$_2$HgSe$_3$ is robust until 50 GPa. However, when the pressure increases up to 54.2 GPa, a set of new peaks emerges and dominates on further compression, indicating the occurrence of a structural phase transition. It should be noted that the superconductivity is observed beyond this pressure.

The pressure-induced structure evolution of Pt$_2$HgSe$_3$ was also confirmed by *in situ* Raman spectroscopy measurements. According to group theory analysis, there are seven Raman-active modes ($3A_{1g}+4E_g$) that can be observed experimentally for Pt$_2$HgSe$_3$[42]. Figure 3c shows the Raman spectra of Pt$_2$HgSe$_3$ at various pressures. The assignments of the modes of Pt$_2$HgSe$_3$ at 1.9 GPa are given as follows: $A_{1g}^3$ = 96.3 cm$^{-1}$, $E_g^3$ = 137.2 cm$^{-1}$, $E_g^2$ = 184.3 cm$^{-1}$, $A_{1g}^2$ = 202.2 cm$^{-1}$, $A_{1g}^1$ = 210.7 cm$^{-1}$, $E_g^1$ = 213.1 cm$^{-1}$. The $E_g^4$ mode has not been observed in ambient condition due to its low scattering efficiency. With increasing pressure, the profile of the spectra remains similar to that at ambient pressure, whereas the observed modes exhibit blue shift, thus showing the normal pressure behavior (Figure 3d). An abrupt disappearance of Raman peaks for pressure near to 50 GPa indicates the structural phase transition to a high-pressure phase. The evolution of the Raman spectra is consistent with our synchrotron XRD patterns. In summary, the Raman study provides further evidence for pressure-induced structural phase transitions.

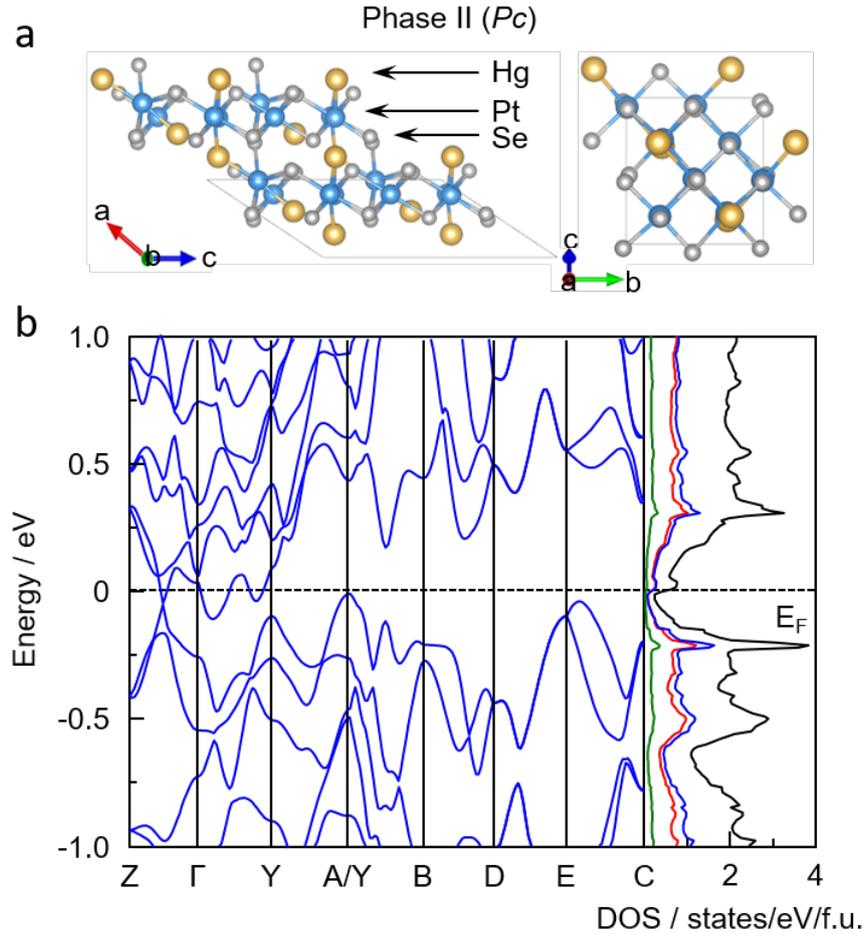

Figure 4. (a) Crystal structure of Pt$_2$HgSe$_3$ in phase II (*Pc*); (b) Electronic band structure and density of states of Pt$_2$HgSe$_3$ at 68 GPa. Black, blue, red and olive line stands for the DOS of whole, Pt, Se and Hg, respectively.

**Pressure induced phase transition and high-pressure structure**

To identify the high-pressure phase, we performed extensive structure searches of Pt$_2$HgSe$_3$ by using our developed structure search method[33, 34]. Structure searches are carried out with simulation cells ranging from one to four formula units and successfully predicted a stable phase (space group *Pc*, Phase II) at 68 GPa. The enthalpy difference curves for the predicted phases are shown in Figure S3a. The Phase II has lower enthalpy than that of the ambient phase (space group $P\bar{3}m1$, Phase I)[14], when the pressure is above 45 GPa, indicating the energetic stability of the newly predicted phase. The experimentally observed phase transition is in apparent agreement with our theoretical prediction. Dynamical structural stabilities of the predicted structure were further investigated by calculating phonon dispersion curves. As shown in Figure S3b, no imaginary frequency was found for the two structures, indicating dynamical stability of predicted structures.

We emphasize that by only relying on the experimental data, the structural evolution of the high-pressure phase is not possible because the XRD peaks are rather weak and broad. However, we have the predicted structure at hand, allowing us to refine the observed XRD data from 58.0 to 96.3 GPa by using the predicted structures. It is remarkable that the uses of the predicted structures gave excellent Rietveld fittings, therefore leading to the unambiguous determination of the high pressure as the predicted structure. The high-pressure phase (Phase II) of $Pt_2HgSe_3$ possesses a monoclinic structure with space group *Pc*. The details of crystallographic data are shown in Table 1. The predicted crystal structure of $Pt_2HgSe_3$ at 68 GPa is shown in Figures 4a. Accompanying the structure transition, the bonding feature changes dramatically. In the Phase II, Hg exists as a dimer with Hg-Hg distance of 2.52 Å, which is much smaller than that shortest value (4.5 Å) in ambient phase. In addition, compared with Phase I, the Pt-Pt, Se-Se distances decrease from 3.67, 3.37 Å to 2.81, 2.54-2.77 Å, respectively, while the Hg-Pt, Hg-Se and Pt-Se distances become diverse.

**Electronic structure and topological properties under high pressure**

To theoretically understand the evolution of the topology under pressures, we have performed DFT calculations of $Pt_2HgSe_3$ in $P\bar{3}m1$ phase. Figure 5a shows the bulk electronic structure of $Pt_2HgSe_3$ at ten different pressures. At all pressures, $Pt_2HgSe_3$ are metallic with electron pockets at K and H, and hole pockets between A-L, A-H, K-H, etc. With the increase of the applied pressure, the electron pockets become larger with the Dirac point at K and H moving to higher binding energy. Meanwhile, the hole pockets between A-L, A-H become larger as well by extending to higher energy. However, the hole pocket between K-H becomes smaller.

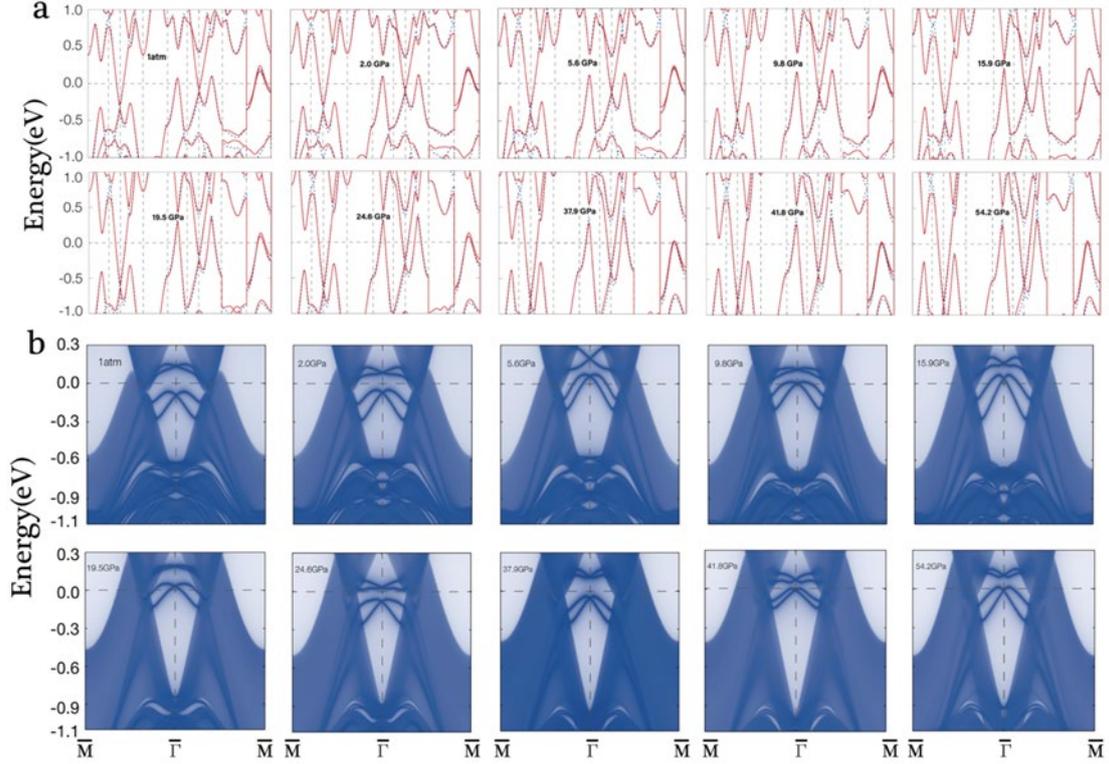

Figure 5. (a) Electronic structure of $Pt_2HgSe_3$ in space group $P\bar{3}m1$ phase at different pressures. The blue dashed lines and red solid lines correspond to the calculations without/with SOC, respectively; (b) The corresponding states at (001) surface calculated from wannier tight-binding model.

We further calculated the electronic states at (001) surface of $Pt_2HgSe_3$ to gain insight into the pressure influence on the topological nature, shown in Figure 5b. We used the selected columns of the density matrix (SCDM) method to automatically generate the Wannier orbitals, which were subsequently used to construct the tight-binding model to reproduce the 144 Bloch bands around the Fermi level ($E_F$) and further calculated the surface states with the iterative Green's function approach[43]. As shown in Figure 5b, at all pressures studied where the $P\bar{3}m1$ structure is preserved, the topological nature of $Pt_2HgSe_3$ is unaffected by the external pressure. Both surface states locating above and below the Fermi level survive at all ten pressures, indicating a robust topological nature of the system, confirming the topology is crystalline symmetry protected.

We further calculated the electronic band structure and the density of states for the newly predicted phase. Figure 4b indicates typical metallic feature of phase II (*Pc* phase) at 68 GPa, with significant contributions from Pt $t_{2g}$ and Se $p_y$ orbitals around the Fermi level. Our results demonstrate that superconductivity observed here is come from phase II. This is obvious different from previous prediction where unconventional superconductivity in monolayer jacutingaite could induce at van Hove filling for electron and hole doping[19]. We also characterized the phase II in terms of the topology, and we found it is trivial. The transition from $P\bar{3}m1$ to $Pc$ phase not only induces the sharp change of resistivity, but also accompanies the topological phase transition originating from the loss of mirror protection present in $P\bar{3}m1$.

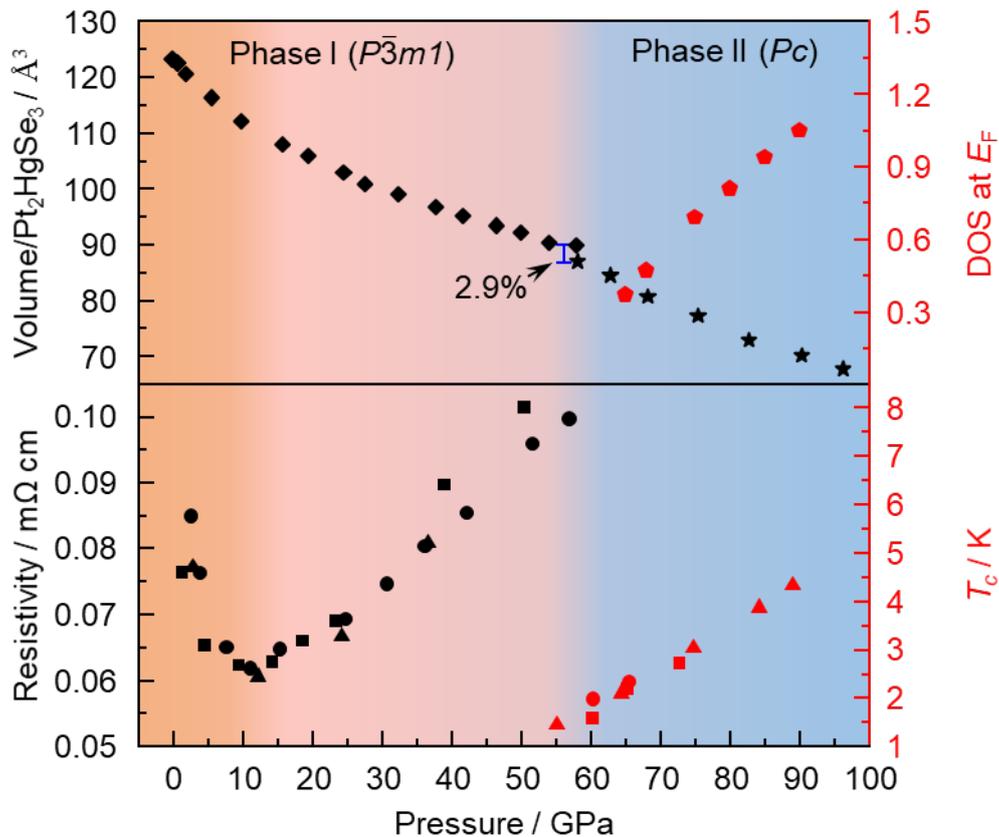

Figure 6. Phase diagram for $Pt_2HgSe_3$. The upper panel shows the pressure dependence of the lattice volume calculated from experiments. The density of states (DOS) at the Fermi level for phase II also shown here; Black rhombus and star stand for lattice volume in phase I and phase II, respectively. Red pentagon stands for DOS at the Fermi level for phase II; the lower panel shows the superconducting $T_c$ as function of pressure and resistivity at 1.8 K for $Pt_2HgSe_3$ in different runs. The values of $T_c$ onset were determined from the high-pressure resistivity. Dot, diamond and triangle stands for $\rho(T)$ measurements in run I, run II and run III, respectively. Black and red stands for resistivity at 1.8 K and $T_c$, respectively.

**Phase diagram of $Pt_2HgSe_3$**

The pressure dependence of the resistivity at 1.8 K and the critical temperature of superconductivity for $Pt_2HgSe_3$ are summarized in Figure 6. It is seen that the high pressure dramatically alters the electronic properties in $Pt_2HgSe_3$. The resistivity first decreases with pressure to a minimum at approximate 11 GPa and then displays the opposite trend with further increasing pressure. The non-monotonic evolution of $\rho(T)$ is also observed in other topological materials[25, 44, 45]. Here, we emphasize that this peculiar behavior of the resistivity in $Pt_2HgSe_3$ is not associated with a structure phase transition based on *in situ* high-pressure XRD measurements. However, the lattice parameter ratio *c*/*a* presents a discontinuous trend at the same pressure, indicating the anisotropy is changed at this pressure (Figure 3b). Our calculations clearly indicate that the topological phase transition cannot explain the resistivity change in $Pt_2HgSe_3$, as the topological nature persists to the structure change at 60 GPa. The evolution of the surface states is strongly affected by the surface potential, which does not monotonically change with the increase of pressure. Thus, we do not observe a uniform behavior of the surface state under the evolution of the pressure. The surface states only take a smaller weight as compared to the bulk states in contribution to the fermi surface. The overall fermi surface enlarges with the increase of pressure. This is consistent with the resistivity measurement at pressures smaller than 11 GPa. However, when pressure is greater than 11 GPa, the experimental resistivity displays an upper turn and further increases with pressure, which cannot be solely explained by the electronic structure predicted by DFT. Meanwhile, the Raman spectra shows that $A_{1g}^2$ mode becomes weak and $E_g^4$ mode emerged at around 10.7 GPa (Figure 3d). Thus, other mechanisms, including the intrinsic electron-phonon interactions, extrinsic defects/vacancies reaction to external pressure, are more likely responsible for this interesting behavior.

At a further increase of pressure, structural phase transition observed accompanying 2.9% volume drops at a critical pressure of 54.2 GPa (as shown on the upper panel of Figure 6). It is clearly seen that the superconducting state emerges beyond the phase transition, and then the superconducting transition temperature increases further with applied pressure. The $T_c$ of $Pt_2HgSe_3$ rises to 4.4 K at the pressure of 88 GPa and still does not exhibit the trend of saturation. As superconductivity occurs among electronic states at $E_F$, we investigated the density of states (DOS) at $E_F$ in the Phase II for various pressures. The upper panel of Figure 6 also shows DOS at the $E_F$ from 65 to 90 GPa. It is clear that the DOS decreases monotonically with increasing pressure. This pressure dependence of the DOS agrees well with the variation of $T_c$.

**CONCLUSIONS**

In conclusion, the evolution of the electrical transport properties in QSHI $Pt_2HgSe_3$ is investigated under high pressure. A non-monotonic evolution of $\rho(T)$ is observed under high pressure. The nontrivial topology in $Pt_2HgSe_3$ is robust and persists to around 55 GPa. The appearance of superconductivity is accompanied by a structural phase transition. Considering effectively tunable of electronic properties and crystal structure in this novel QSHI, $Pt_2HgSe_3$ offers a new platform for exploring exotic physics upon compression.

**Notes**
The authors declare no competing financial interests.

**ACKNOWLEDGMENT**

We thank Prof. Yanming Ma for valuable discussions. This work was supported by the National Key R&D Program of China (Grant No. 2018YFA0704300 and 2017YFE0131300), the National Natural Science Foundation of China (Grant No. U1932217, 11974246, 11874263, and 12074138), the Natural Science Foundation of Shanghai (Grant No. 19ZR1477300) and the Science and Technology Commission of Shanghai Municipality (19JC1413900). P. N-I. acknowledges support from the Hungarian Academy of Sciences, Lendület Program, grant no. LP2017-9/201. A.V. acknowledges the support from the Czech Geological Survey (DKRVO/CGS 2018–2022). The authors thank the support from Analytical Instrumentation Center (# SPST-AIC10112914), SPST, ShanghaiTech University. We thank the staffs from BL15U1 at Shanghai Synchrotron Radiation Facility, for assistance during data collection. We used the computing facility at the High-Performance Computing Centre of Jilin University.

**Table 1.** Structural parameters of Pt$_2$HgSe$_3$ under different pressures at room temperature.

|  | 1 atm | 82.7 GPa | |
|---|---|---|---|
| Phase | Phase I | Phase I | Phase II |
| Crystal system | trigonal | trigonal | monoclinic |
| Space group | $P\bar{3}m1$ (164) | $P\bar{3}m1$ (164) | $Pc$ (7) |
| a | 7.341132(2) | 6.781756(4) | 7.166096(1) |
| b | 7.341132(2) | 6.781856(4) | 6.045144(3) |
| c | 5.275207(3) | 4.004362(2) | 12.129331(3) |
| α | 90 | 90 | 90 |
| β | 90 | 90 | 146.262 |
| γ | 120 | 120 | 90 |
| atoms position | Wyckoff (x y z) | Wyckoff (x y z) | Wyckoff (x y z) |
| Pt1 | 1a (0,0,0) | 1a (0,0,0) | 2a (0.8182,-0.0063,0.4345) |
| Pt2 | 3e (0.5,0,0) | 3e (0.5,0,0) | 2a (0.8042,-0.5047,0.4446) |
| Hg1 | 2d (1/3,2/3,0.3513) | 2d (1/3,2/3,0.3507) | 2a (-0.4053,-0.2606,0.1690) |
| Se1 | 6i (0.8196,0.1804,0.2492) | 6i (0.8196,0.1804,0.2492) | 2a (1.0939,-0.4986,0.9235) |
| Pt3 |  |  | 2a (0.8121,-0.2448,0.6904) |
| Pt4 |  |  | 2a (0.9924,-0.7349,0.7145) |
| Hg2 |  |  | 2a (1.2971,-0.7646,0.6947) |
| Se2 |  |  | 2a (1.1268,-0.9905,0.9236) |
| Se3 |  |  | 2a (0.5572,-0.9779,0.4716) |
| Se4 |  |  | 2a (1.1155,-0.2482,0.6763) |
| Se5 |  |  | 2a (0.5097,-0.2511,0.7051) |
| Se6 |  |  | 2a (1.4820,-0.5166,0.9421) |
| Residuals[a] / % | $R_{wp}$: 2.82 | $R_{wp}$: 1.32 | |
|  | $R_p$: 1.88 | $R_p$: 0.97 | |

[a]Here $R_p = \Sigma||F_{obs}|-|F_{calc}||/\Sigma|F_{obs}|$ and $R_{wp} = (\Sigma[w(|F_{obs}|^2-|F_{calc}|^2)^2]/\Sigma[w(|F_{obs}|^2)^2])^{1/2}$, where $F_{obs}$ is the observed structure factor and $F_{calc}$ is the calculated structure factor.

.